\pdfoutput=1
\documentclass[twocolumn,english,aps,superscriptaddress,prb,longbibliography,10pt]{revtex4-1}

\usepackage{babel}
\usepackage{amsmath}
\usepackage{amssymb}
\usepackage{graphicx}
\usepackage{physics}

\usepackage{comment}
\usepackage{xcolor}

\usepackage[
  colorlinks=true,
  urlcolor=blue,
  linkcolor=blue,
  citecolor=blue,
]{hyperref} 

\hypersetup{
 	pdftitle={Thermal Phase Diagram of the Square Lattice Ferro-antiferromagnetic J1-J2 Heisenberg Model},
	pdfauthor={Olivier Gauthé and Frédéric Mila},
}

\begin{document}

\title{\texorpdfstring{Thermal Phase Diagram of the Square Lattice Ferro-antiferromagnetic $J_1-J_2$ Heisenberg Model}{Thermal Phase Diagram of the Square Lattice Ferro-antiferromagnetic J1-J2 Heisenberg Model}}

\author{Olivier Gauth\'e}
\affiliation{Center for Computational Quantum Physics, Flatiron Institute, 162 5th Avenue, New York, NY 10010.}
\affiliation{Institute of Physics, \'Ecole Polytechnique F\'ed\'erale de Lausanne (EPFL), CH-1015 Lausanne, Switzerland}
\author{Fr\'ed\'eric Mila}
\affiliation{Institute of Physics, \'Ecole Polytechnique F\'ed\'erale de Lausanne (EPFL), CH-1015 Lausanne, Switzerland}

\date{\today}
\begin{abstract} 
Using state of the art tensor network computations combined with spin wave theory, we compute the finite temperature phase diagram of the spin 1/2 $J_1-J_2$ Heisenberg model on the square lattice with ferromagnetic $J_1 < 0$ and antiferromagnetic $J_2 > 0$. We localize the first order line originating from the first order zero temperature point as well as the second order thermal transition at large $J_2$ associated with spontaneous lattice rotation symmetry breaking. This second order line terminates at a critical end point where it meets the first order line. We found no evidence for an intermediate phase.
\end{abstract}

\maketitle

\section{Introduction}

The antiferromagnetic $J_1-J_2$ Heisenberg model for spin-1/2 is a cornerstone model of frustrated magnetism~\cite{chandra_possible_1988} and has been extensively studied. The ferromagnetic case with $J_1 < 0$ is also of interest, both for theoretical and experimental motivations.
A conceptually simple model with only one tuning parameter, it nonetheless displays surprisingly rich physics with several phase transitions driven by strong quantum fluctuations.
Its study has been motivated by the possibility for quantum fluctuations to stabilize an exotic intermediate phase around the maximally frustrated point $J_2/J_1 = 0.5$, in addition to the ferromagnetic and collinear antiferromagnetic phases of the classical version. Beyond its theoretical relevance, experimental realizations
have been proposed in vanadium phosphates~\cite{kaul_evidence_2004,tsirlin_extension_2009, janson_magnetic_2016, bhartiya_inelastic_2021, landolt_spin_2022}.

Investigating it at either zero or finite temperature proves challenging due to the competing nature of interactions, very small energy differences between phases and large finite size effects. Numerical studies struggle due to the infamous sign problem that prevents the use of quantum Monte Carlo in most cases. Thanks to advances in tensor network methods, this model is now open to simulation and offers a rich playground to study phase transitions and their intersections.

Here we consider the model at finite temperature and investigate the different phases and the associated transitions. The phase diagram is rich, with both first and second order transitions, one critical point and one critical end point. Indeed our data support the second order line to be cut at finite temperature by the first order line, without any intermediate phase.

This article is organized as follows: in section II we introduce the model and the results established in previous studies. In section III we expose our methodology based on tensor network algorithms and spin wave theory, and how we detect the different phase transitions.
In section IV we present our main results and draw the finite temperature phase diagram of the model. We summarize our results and conclude in section V.

\section{Model}
We study the spin-1/2 Heisenberg model on the square lattice with ferromagnetic first neighbor interaction $J_1 < 0$ and frustrated second neighbor interaction $J_2 > 0$ at finite temperature. We take $J_1 = -1$ as the energy scale and write the Hamiltonian

\begin{equation}
\mathcal{H} = -\sum_{\expval{i,j}} \textbf{S}_i \cdot \textbf{S}_j
+ J_2 \sum_{\langle \expval{i,j} \rangle} \textbf{S}_i \cdot \textbf{S}_j
\end{equation}

For classical spins, this model is mapped to the antiferromagnetic model with $J_1 > 0$ by a $\pi$ rotation of every other site. Consequently the classical phase diagram is the same in both cases, with a zero-temperature phase transition at $J_2 = \abs{J_1}/2$ and a stripe phase breaking lattice rotation at finite temperature in the large $J_2$ region~\cite{chandra_ising_1990,weber_ising_2003,shannon_finite_2004}.

In the case of spins-1/2, it is known that quantum fluctuations play a major role in the maximally frustrated region and can generate a totally different physical behavior. In the antiferromagnetic model, they give birth to at least one intermediate phase that does not break SU(2) symmetry, whose full characterization has been the subject of intense debates~\cite{gelfand_zero-temperature_1989, read_valence-bond_1989, dagotto_phase_1989, figueirido_exact_1990, bishop_phase_1998, jiang_spin_2012, hu_direct_2013, gong_plaquette_2014, poilblanc_quantum_2017, haghshenas_u1-symmetric_2018, yu_deconfinement_2018, wang_critical_2018, liu_gapless_2018, hasik_investigation_2021,ferrari_gapless_2020, liu_gapless_2022, nomura_dirac-type_2021}. The ferromagnetic case is also of interest and has been the subject of both analytical and numerical studies. However, despite the exact mapping between classical versions, the ferromagnetic model is surprisingly different from the $J_1 > 0$ case. Quantum fluctuations still strongly impact the physics of the model, with a first order transition at finite temperature absent from the classical case as we shall see, but they have different outcomes.

In the low $J_2$ region, the system is dominated by the ferromagnetic interaction and the exact ground state is the saturated ferromagnetic state $\ket{\uparrow \uparrow \dots \uparrow}$ (more generally any state belonging to the maximal spin multiplet). There are no quantum fluctuations, this ground state is the same as for classical spins and the exact energy per site is $E_{\text{FM}} = (-1 + J_2)/2.$
At finite temperature, quantum fluctuations may slightly modify physical quantities without changing the main picture of a classical ferromagnet. Due to Mermin-Wagner theorem, the system spontaneously breaks SU(2) symmetry at zero temperature only, at finite temperature no symmetry is broken.

In the large $J_2$ region, the system orders as a collinear antiferromagnet to fulfill the second neighbor interaction. The collinear Néel ground state spontaneously breaks both spin SU(2) and lattice symmetries, with order by disorder selecting magnetic stripes in either vertical or horizontal direction.

The nature of the ground state in the intermediate region and the possibility of an intermediate phase have been discussed for many years. The exact ferromagnetic energy being known, its value can be used to look for an instability of the ferromagnetic state and deduce a critical value $J_2^c$ for the second neighbor interaction. One magnon calculation gives the classical value $J_2^c = 0.5$ while two-magnon gives $J_2^c = 0.4082$~\cite{dmitriev_two-dimensional_1997}.
A  variational wavefunction approach sets a definitive upper bond $J_2^c \leq 0.4056$~\cite{dmitriev_two-dimensional_1997}. 
Earlier numerical works based on exact diagonalization~\cite{shannon_nematic_2006}, projected wavefunctions~\cite{shindou_projective_2011} and pseudofermion functional
renormalization group~\cite{iqbal_intertwined_2016} proposed the existence of a narrow spin nematic phase breaking SU(2) symmetry in a very subtle way. This spin nematicity would emerge from Bose-Einstein condensation of magnon pairs~\cite{zhitomirsky_magnon_2010}.

However the coupled cluster method~\cite{richter_frustrated_2010} does not detect any intermediate phase, with a transition out of the ferromagnetic phase at $J_2^c = 0.394$ and a non-zero order antiferromagnetic order parameter developing as early as $J_2 = 0.4$. More recent calculations based on cylinder DMRG~\cite{jiang_where_2023} agree with $J_2^c \approx 0.39$ and conclude that the nematic phase is only stabilized under an external field: for the SU(2) symmetric model, there is a direct, first order transition from the ferromagnetic phase to the collinear N\'eel phase. Our own finite temperature data agree with this scenario and we will use the value $J_2^c = 0.394$ as the quantum point.

\begin{figure}[ht!]
\includegraphics[width=1.0\linewidth]{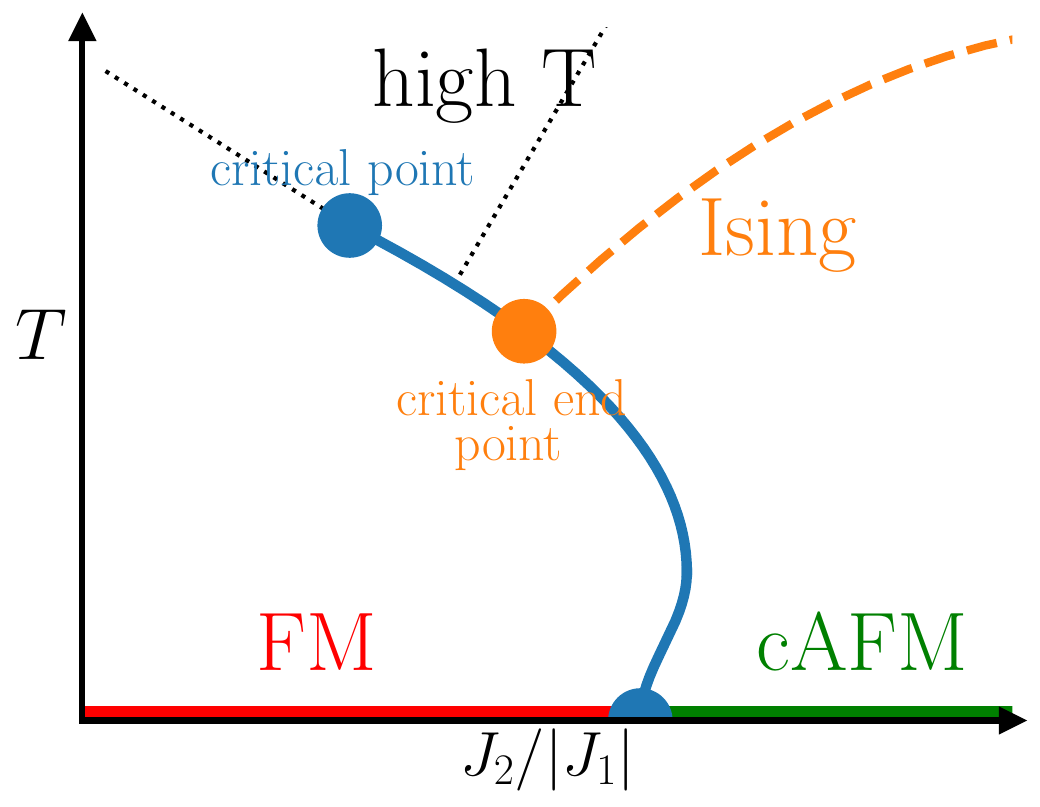}
\caption{\footnotesize{Sketch of the finite temperature phase diagram for the spin-1/2 ferro-antiferromagnetic $J_1-J_2$ model. A first order line emerges from the zero temperature first order point between the ferromagnetic phase (FM) and the collinear N\'eel phase (cAFM) with an infinite slope and terminates at a critical point. Above the collinear antiferromagnet, lattice symmetry is spontaneously broken at low temperature with an Ising transition. This second order line finishes at a critical end point when it meets the first order line. The dotted lines mark the crossovers associated with specific heat maxima.}}
\label{fig:sketch}
\end{figure}

At finite temperature, the first order quantum point naturally extends to a first order line that ends at a critical point, as observed in the Shastry-Sutherland model~\cite{jimenez_quantum_2021}. In the collinear antiferromagnetic phase, lattice rotation symmetry is spontaneously broken by the stripe direction selection. While N\'eel order itself can only appear at zero temperature due to Mermin-Wagner theorem, the spontaneous breaking of the finite group $C_{4v}$ occurs at finite temperature with a phase transition in the Ising universality class~\cite{chandra_ising_1990} similarly to the classical~\cite{weber_ising_2003} and $J_1 > 0$~\cite{gauthe_thermal_2022} cases.
In this work, we provide numerical evidence showing that this second order line ends at a critical end point when it meets the first order line (see Fig.~\ref{fig:sketch}).

\section{Method}

\subsection{Tensor network algorithms}
We investigate this model using tensor network methods, specifically Projected Entangled Pair States (PEPS)~\cite{verstraete_renormalization_2004}. We use the framework of purification to construct an approximation of the finite temperature density matrix at thermal equilibrium~\cite{verstraete_matrix_2004}. In this setup, each physical variable on the lattice with dimension $d=2$ is described by a local tensor also carrying another auxiliary spin-1/2 called ancilla that acts as a thermal bath. Correlations between different sites are carried through virtual variables with a bond dimension $D$ associated with the maximal amount of entanglement allowed in the purified wavefunction.

We start from the exact infinite temperature state, where each spin is maximally entangled with its ancilla partner and the purified wavefunction is a dimer product state. We then use second neighbor simple update~\cite{jiang_accurate_2008, corboz_simulation_2010} to imaginary time evolve up to the desired temperature with a finite imaginary time step $\tau$. Independently from time evolution, we use the Corner Transfer Matrix Renormalization Group (CTMRG) algorithm to  approximately contract the tensor network~\cite{nishino_corner_1996, orus_simulation_2009} using the directional algorithm~\cite{corboz_stripes_2011, corboz_competing_2014}. The corner dimension $\chi$ controls the contraction accuracy.
A similar method has already been used successfully to investigate the critical point in the Shastry-Sutherland model~\cite{jimenez_quantum_2021} as well as the second order transition line in the antiferromagnetic case~\cite{gauthe_thermal_2022} (see Ref.~\cite{gauthe_thermal_2022}  and its supplemental material for a detailed discussion of the algorithm).
We implement SU(2) symmetry at the tensor level for simple update~\cite{singh_tensor_2012, schmoll_programming_2020}, however performance limitations prevents us from using it for the much more expensive CTMRG part. There, we improved the U(1) code that we previously used~\cite{gauthe_thermal_2022} by combining $S^z$ conservation with the non-commuting $S^z$ reversal symmetry, resulting in O(2) symmetry implementation with substantial performance gains (see appendix \ref{ap:sym}).

Once the CTMRG environment is converged, local observables are easily computed. The specific heat is obtained by numerically differentiating the energy. The transfer matrices in both horizontal and vertical directions are approximated using the environment tensors, their eigenvalues $\lambda_i$ directly provide correlation lengths as $1/\xi_i = \ln{\abs{\lambda_i / \lambda_0}}$ without the need to compute long distance correlation functions.

We computed most of our results with a bond dimension $D=19$. In the ferromagnetic phase, the bond dimension has very little impact on the result and the corner dimension $\chi$ only affects the measured correlation length. In the large $J_2$ region, finite $D$ effects are stronger, which may affect the Ising transition temperature. Changing the imaginary time step $\tau$ impacts the ability to converge the CTMRG by modifying the magnitude of the  simple update induced $C_{4v}$ asymmetry,  however whenever convergence can be reached the effects of $\tau$ on short distance observables are negligible (see appendix~\ref{ap:control} for more information on simulation parameter control).

\subsection{Modified spin wave theory}
In the ferromagnetic region, we use modified spin wave theory (MSWT) to obtain an expression for the free energy at finite temperature. This semi-classical method builds on the exact ferromagnetic ground state and adds both thermal and quantum fluctuations with a chemical potential that enforces a zero magnetization~\cite{takahashi_quantum_1986, takahashi_few-dimensional_1987}. We present here the main elements and provide a detailed derivation in appendix~\ref{ap:mswt}.

Starting from the one-magnon energy Eq.~(\ref{eq:epsk}),
\begin{widetext}
\begin{equation}
\epsilon_\textbf{k} = 2\left(\sin^2\frac{k_x}{2} + \sin^2\frac{k_y}{2}\right)  - 2 J_2\left(\sin^2\frac{k_x + k_y} {2} + \sin^2\frac{k_x-k_y}{2}\right)
\label{eq:epsk}
\end{equation}
\end{widetext}
the chemical potential $\mu$ is obtained at inverse temperature $\beta$ by (numerically) solving the implicit equation defined by an integral over the Brillouin zone
\begin{equation}
\frac{1}{2\pi^2}\iint_{\text{BZ}} \frac{\dd^2 \vb{k}}{\exp(\beta(\epsilon_{\vb{k}} - \mu)) - 1} - 1 = 0
\label{eq:mu}
\end{equation}
and results in an expression for the free energy
\begin{equation}
f = E_{\text{FM}} + \frac{\mu}{2} - \frac{1}{4\pi^2\beta}\iint_{\text{BZ}} \ln(1 + \frac{1}{e^{\beta(\epsilon_{\vb{k}} - \mu)}-1})\dd^2 \vb{k}.
\end{equation}

A similar approach could be used in the large $J_2$ region, however there the ground state differs from the classical collinear N\'eel state: we expect quantum fluctuations to be too strong to obtain quantitatively precise results.  The qualitative behavior of the first order line in the limit $T \rightarrow 0$ can still be obtained from general arguments on the free energy.
Inside a given phase, all thermodynamic quantities are smooth as a function of $T$ and $J_2$. Phases on both sides of the transition are gapless with long-range physics governed by massless Goldstone modes, however they obey different low temperature scaling laws. In the ferromagnetic phase, the free energy is quadratic, associated with quadratic dispersion in the spin wave energy (\ref{eq:epsk}). A Talyor expansion for the free energy in the ferromagnetic phase close to the quantum point writes $F_{\text{FM}}(T) = E_0 + \alpha_1(J_2 - J_2^c) - bT^2 + O(T^3) + O((J_2-J_2^c)^2)$, where $E_0$ is the ground state energy at the transition, $\alpha_1 > 0$ and $b > 0$. In the collinear antiferromagnetic phase, the ground state energy will still be linear in $J_2 - J_2^c$ at first order, but the dispersion relation will be linear in $\textbf{k}$ and generate a $O(T^3)$ temperature dependence in the free energy. The free energy expansion is $F_{\text{AF}}(T) = E_0 - \alpha_2(J_2 - J_2^c) + O(T^3) + O((J_2-J_2^c)^2)$, with $ \alpha_2 > 0$.
 Imposing the free energies to become equal at the transition yields the curve
\begin{equation}
T = \alpha \sqrt{J_2 - J_2^c},
\label{eq:sqJ2}
\end{equation}
that is an infinite slope at $T=0$ and a curve bending towards \emph{larger} $J_2$. Since we observe no trace of this behavior in our numerical data, but rather a turn towards \emph{smaller} $J_2$ at larger temperature, we think this right bend should have very limited $J_2$ extension. 
A similar change of direction has already been observed in the case of the frustrated bilayer (see figure S3 in the supplemental material for Ref.~\cite{stapmanns_thermal_2018}).

\subsection{Second order transition temperature extrapolation}
\begin{figure}
\centering
\includegraphics[width=1.0\linewidth]{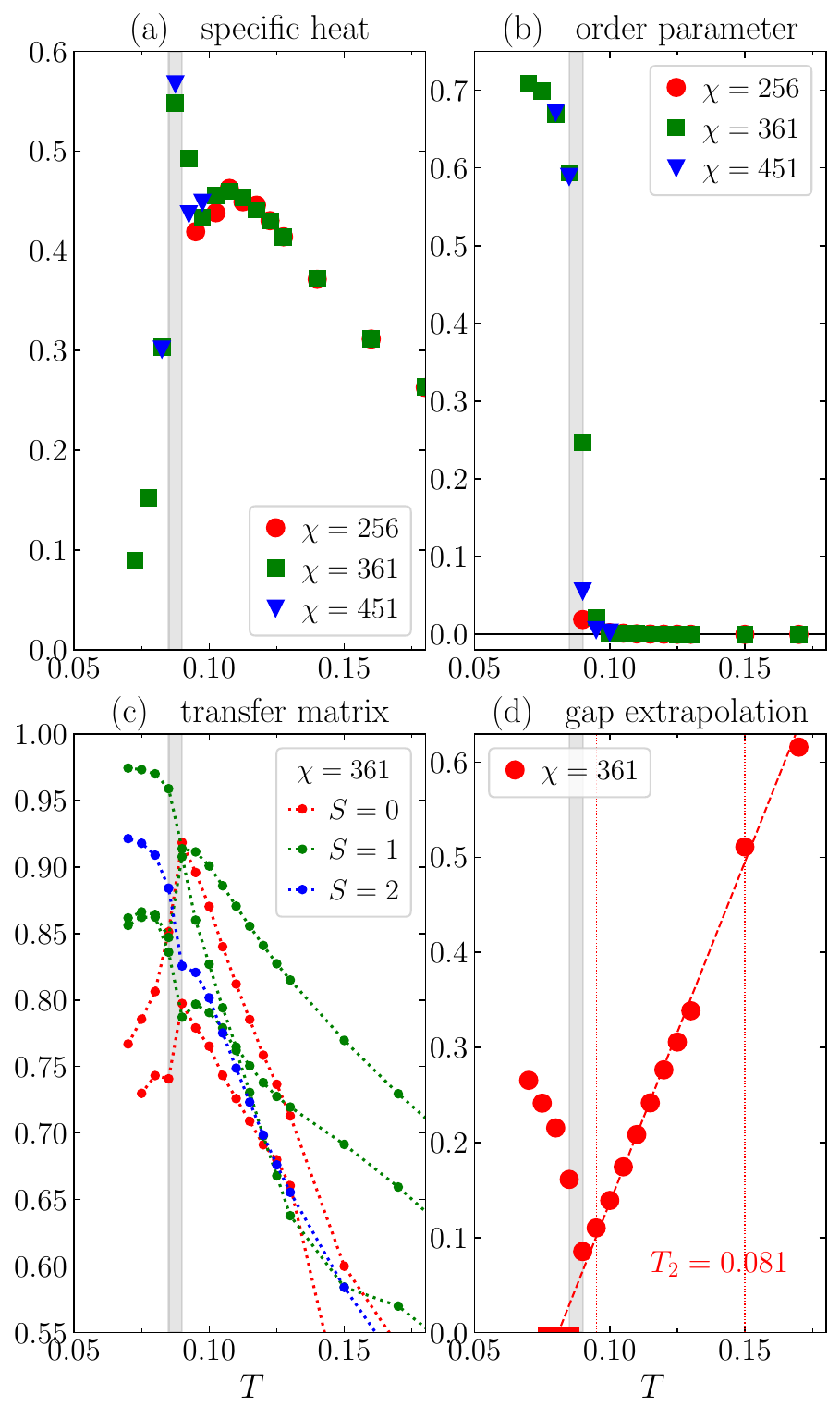}
\caption{\footnotesize{Several independent methods are used to evaluate the Ising critical temperature $T_2$ at $J_2/\abs{J_1} = 0.50$ with $D=19$ and return the same value. (a) The specific heat has a first Schotkky local maximum, then a very narrow, higher peak. (b) The order parameter develops a non-zero value. (c) The transfer matrix spectrum at $\chi=361$ has an accident visible in all spin symmetry sectors. These methods give a critical temperature $0.085 \leq T_2 \leq 0.090$ (grey area). (d) The leading singlet gap is linearly extrapolated, with the red lines marking the fit range. This last method is indirect and less precise but allows to estimate $T_2$ in cases where tensor contraction fails before $T_2$ is reached.}}
\label{fig:t2_extra}
\end{figure}

For large values of $J_2$, the ground state belongs to the N\'eel collinear phase. Stripes appear at finite temperature with a second order phase transition in the 2D Ising universality class and associated with a $C_{4v}$ symmetry breaking order parameter. As we showed  in the $J_1 > 0$ case~\cite{gauthe_thermal_2022}, several converging approaches provide an estimation for the critical temperature $T_2$. We follow a similar analysis here (see Fig.~\ref{fig:t2_extra}). First, a very narrow specific heat peak is spotted, corresponding to a vertical slope in the energy. Second, a non-normalized order parameter $\sigma$ can be defined as the bond energy difference between horizontal and vertical bonds over the unit cell:
\begin{equation}
\sigma = \sum_{\ev{i,j},v} \textbf{S}_i \cdot  \textbf{S}_j - \sum_{\ev{i,j},h} \textbf{S}_i \cdot  \textbf{S}_j
\label{eq:sigma} 
\end{equation}
Due to finite $\chi$ effects, $\sigma$ is smoothed and not strictly vertical at the transition point. Looking at different $\chi$ values, its value flows to 0 for $T > T_2$ and to a finite value for $T< T_2$. We can also construct an approximation of the transfer matrix along either the horizontal or vertical direction and compute its spectrum. The eigenvalues are sorted according to spin multiplet and normalized with respect to a leading eigenvalue that is set to 1.  Close to the transition, the leading excitation in the transfer matrix is also a spin singlet and  the transition itself corresponds to a vanishing singlet gap. While finite $\chi$ effects keep this gap finite in practice, the second leading eigenvalue becomes a singlet only in a narrow region around the transition and then an accident can be seen in all symmetry sectors.

These three methods provide clear, non-ambiguous and mutually agreeing signals for a second order transition at finite temperature to a $C_{4v}$ symmetry breaking phase. They however require data points at the transition itself, where converging the CTMRG is hard. 
Worse, in the intermediate region, the CTMRG does not converge at all at low temperature. This seems to be associated with the multi-site unit cell: the simple update explicitly breaks all lattice symmetries with an error $O(\tau)$, including  translation. As corner singular spectra become nearly continuous close to a second order line or a critical point, slightly different truncations are made on different unit cell site environments, leading to inconsistent global environment and failure to converge. This effect can be reduced using smaller imaginary time step, which reduces the asymmetry between the different sites. Local observables are unaffected by this change in $\tau$. Still, for intermediate values of $J_2$, we did not manage to converge the CTMRG algorithm at low temperature and the Ising phase transition cannot be directly observed.
Fortunately, we can still extrapolate the leading singlet gap of the transfer matrix: it has to linearly vanish at  the critical temperature. This method is less precise and more sensitive to finite $\chi$ effects, but when it can be compared to the previous ones its results are in reasonable agreement.

\subsection{First order temperature extrapolation}

\begin{figure}
\centering
\includegraphics[width=1.0\linewidth]{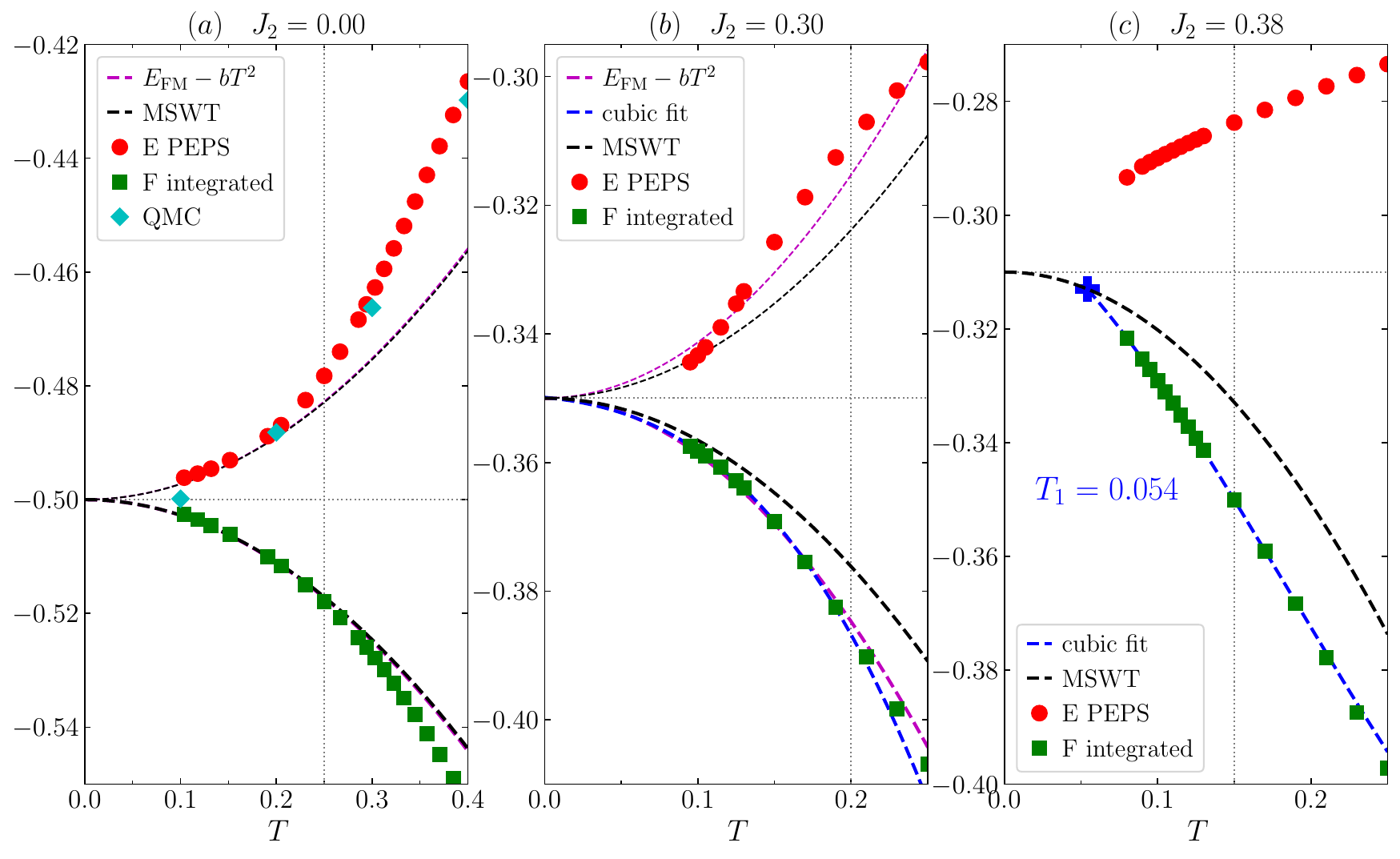}
\caption{\footnotesize{Modified spin wave theory (MSWT) is compared to PEPS with $D=19$ to estimate the first order transition temperature $T_1$ (see text). The free energy obtained 
either from MSWT (black)
or by integrating PEPS energies (green) is fitted and
provides a Taylor expansion for both the free energy (below ferromagnetic energy, thick) and the energy (above, thin). The horizontal (resp. vertical) grey dashed line labels the ferromagnetic energy (resp. the fit range). 
(a) The unfrustrated $J_2 = 0$ case matches Quantum Monte Carlo (QMC) on a $16\times 16$ lattice from Ref.~\cite{okabe_monte_1988} (we consider the very last QMC point to be off due to finite size effects). At low temperature, a quadratic fit with imposed ferromagnetic ground state energy perfectly matches both PEPS and MSWT. (b) At $J_2 = 0.30$, MSWT and PEPS are still compatible at low enough temperature. A cubic fit of the free energy yields $E_0 = E_{\text{FM}}$ and is compatible with a ferromagnetic expansion $F(T) = E_{\text{FM}} + bT^2$: there is no first order transition.  (c) at $J_2 = 0.38$, PEPS data are not compatible with MSWT as the cubic fit gives a higher $T=0$ energy than the ferromagnetic ground state: fitted points do not belong to the ferromagnetic phase. The crossing point between the free energies provides an estimation for the transition temperature.}}
\label{fig:mswt}
\end{figure}

We were not able to precisely localize the critical point nor to compute observables close the first order line as the CTMRG fails to converge close to them. To estimate the first order transition temperature, we use an indirect method by comparing PEPS and MSWT free energies. 
The entropy at a given temperature is evaluated from a set of energy points at higher temperature by integrating the specific heat starting from $\beta=0$: after partial integration one gets
$S(\beta) = \ln 2 + \beta  E(\beta) - \int_0^\beta E(\beta') d\beta'.$
This integral can be evaluated numerically and one obtains the PEPS free energy $F=E-TS$.
To extrapolate values beyond the temperature range where CTMRG converges, we rely on fits and Taylor expansions. Fitting directly the energy yields poor fits as its concavity changes at a specific heat maximum, on the other hand the free energy is always concave and behaves in a much nicer way.

In the ferromagnetic phase, the free energy is quadratic with respect to temperature and the exact ground state energy is known: we use this ansatz to fit the MSWT with negligible error to interpolate between computed points. A free energy expansion $F(T) =  E_{\text{FM}} - bT^2$ also provides an expansion for the energy $E(T) = E_{\text{FM}} + bT^2$. We benchmark modified spin wave theory at $J_2 = 0$ in Fig.~\ref{fig:mswt} (a) and find perfect agreement with our numerical data. Hence we consider MSWT to give a reliable free energy value of the ferromagnetic phase at low temperature.

In a 2D antiferromagnet, we expect vanishing linear and quadratic terms with a purely cubic free energy $F(T) = E_0 - cT^3$, however this expansion does not match our points, which are too high in temperature: we cannot expect this behavior in the high temperature phase, above the second order line. As we cannot rely on a free energy expansion at $T=0$, we fit our data with a generic cubic curve: this is similar to writing a Taylor expansion at a temperature $T' > 0$, it implies we need to consider linear and quadratic orders as well. We impose $E_0 \geq E_{\text{FM}}$ as we know from literature the ground state is still ferromagnetic but no other restriction on the parameters. As shown in Fig.~\ref{fig:mswt} (b), for $J_2 = 0.30$, this cubic fit recovers $E_0 = E_{\text{FM}}$ and the linear term is close to zero. The fit is actually very close to the one obtained from the ferromagnetic free energy expansion and is compatible with MSWT at low temperature. 

Everything changes in the case $J_2 = 0.38$ plotted in Fig.~\ref{fig:mswt} (c), where we see a clear discrepancy  between PEPS and MSWT: at intermediate temperature they belong to different phases. The free energy intersection point at $T_1 = 0.054$ gives an estimation for the first order transition temperature, where the two phases have the same free energy. Since the free energy is always concave, a linear extrapolation of the same free energy points always overestimates the free energy and provides an upper bond for the crossing point. Here, a linear fit over the same temperature range provides an upper bound $T_1 < 0.055$, very close to the cubic fit value.

\section{Phase diagram}
\begin{figure}
\centering
\includegraphics[width=1.0\linewidth]{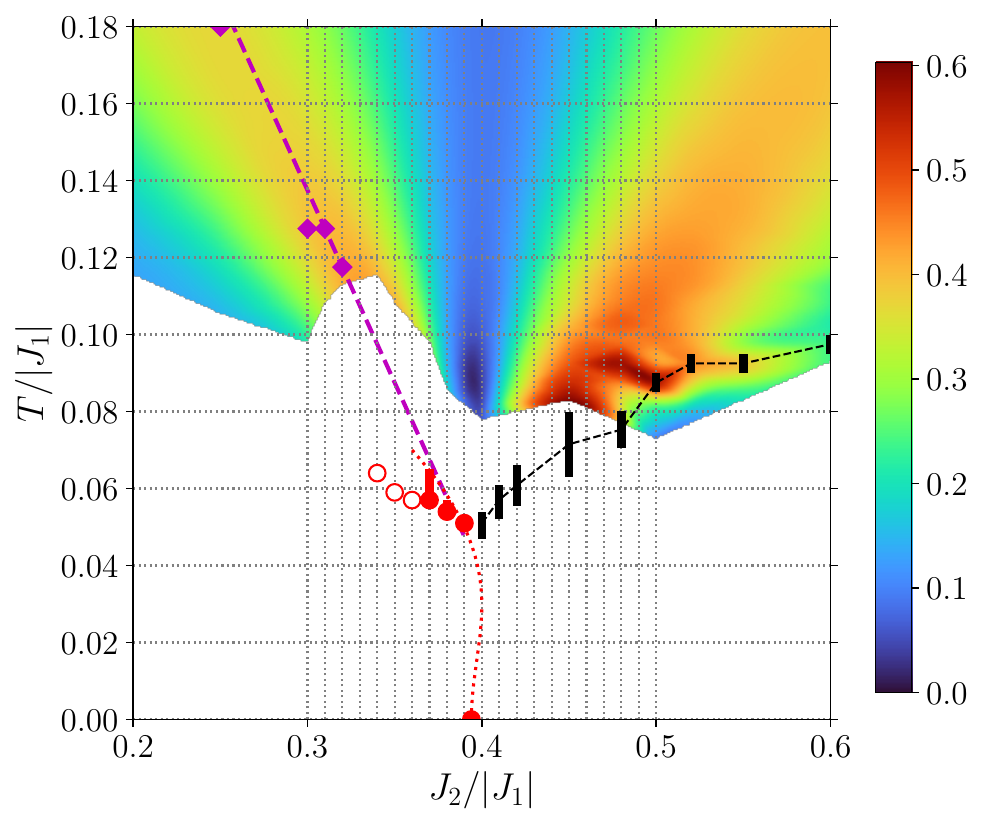}
\caption{\footnotesize{Zoom on the intermediate region. The specific heat maximum position above the ferromagnetic region (diamonds) can be linearly extrapolated (magenta line) to match the first order points obtained from the free energy crossings. It is compatible with the free energy crossing points for $J_2 \geq 0.37$. The empty red points are the crossing points obtained with the same method that we do not trust due to their distance to the magenta line. The Schottky-like broad peak for large $J_2$ is expected to persist for smaller $J_2$ as a separate peak above the Ising transition until it hits the first-order transition line (see text). The gray dotted lines are guides for the eye.}}
\label{fig:zoom}
\end{figure}

We can now combine together PEPS observables and phase boundaries to draw the full phase diagram.
Drawing the second order line for large values of $J_2$ is straightforward using the method explained in Fig.~\ref{fig:t2_extra}. In the intermediate region, we need to compare the first order temperature $T_1$ and the extrapolated second order critical temperature $T_2$. For $J_2 \leq 0.39$, we get $T_1 > T_2$: the second order transition never occurs as the system directly enters the ferromagnetic phase.
For $J_2 = 0.40,$ we find $T_1 < T_2$: in this case there should be an Ising transition to the stripe phase. This transition may be followed by a first order transition to the ferromagnetic phase before a final reentrance to the stripe phase, as we do not know the $J_2$ extend of the first order line. In any case the extrapolation method for $T_1$ is no more valid as the free energy fit in the high temperature phase cannot provide a valid expression for the free energy in the $C_{4v}$ broken phase. We clearly see e.g. on Fig.~\ref{fig:zoom} that the second order temperature does not flow to zero before meeting the first order line: there has to be a critical end point in the phase diagram where the Ising transition line terminates. From the analysis at $J_2 = 0.39$ and $J_2 = 0.40$, we have an estimation for the position of the critical end point with $0.39 < J_{2}^\text{CEP} < 0.40$, and a temperature  $T_\text{CEP}\sim 0.05$.

The maximal $J_2$ value reached by the first-order line cannot be obtained with our current method, although as we mentioned we expect it to be very close to the zero temperature point. To localize the other extremity of the line, we consider the specific heat in the ferromagnetic region (see the zoom in Fig.~\ref{fig:zoom}). Indeed, we understand the specific heat peak to be a smooth continuation of the first order line as a crossover, therefore its position can help us to localize the first order line. The temperature of the specific heat maximum at fixed $J_2$ is close to linear in $J_2$, therefore we can extrapolate it and see when it matches the free energy crossing points. For $J_2 \geq 0.37$, this extrapolation is compatible with the points and their error bar, so we think the method is valid in that region. For $J_2 \leq 0.36$, the crossing point temperature seems too low and we believe there is no transition but a crossover to a ferromagnetic behavior. Accordingly, we estimate the critical point should be close to $J_2 \sim 0.37$.

\begin{figure*}[ht]
\includegraphics[width=1.0\textwidth]{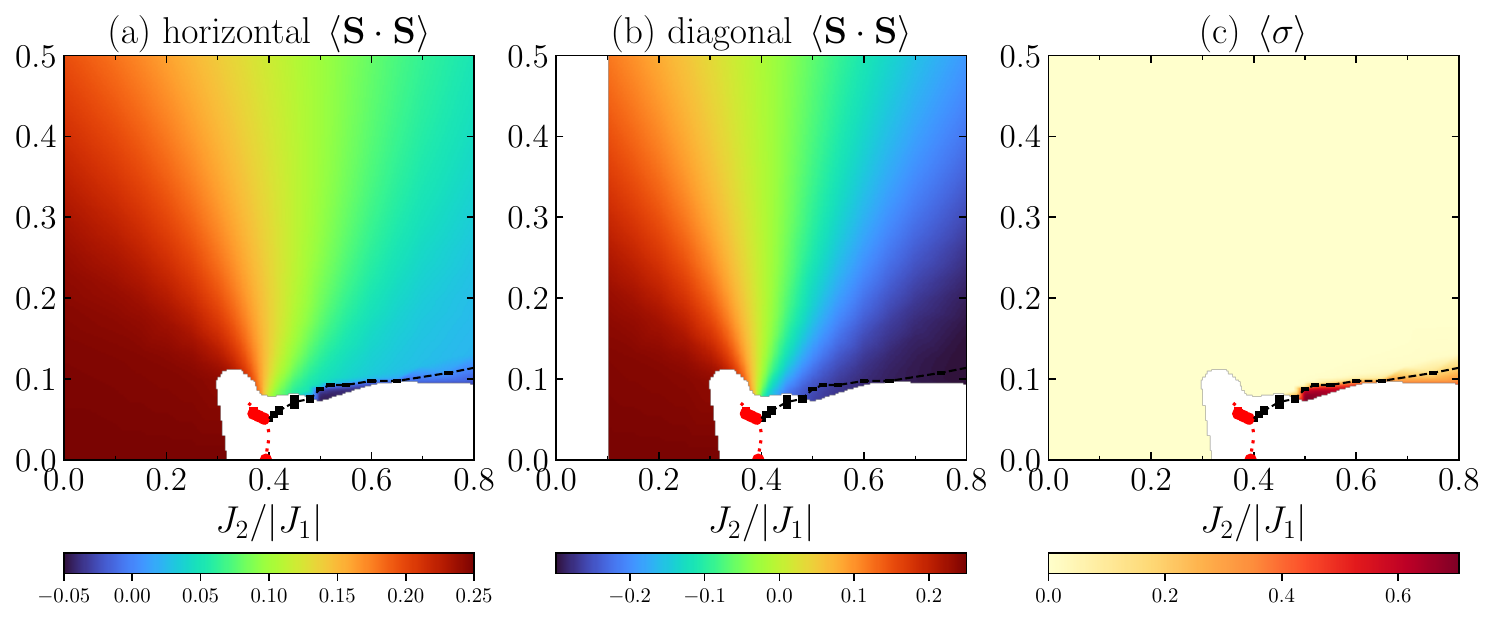}
\caption{\footnotesize{Expectation value for the observable $\textbf{S}\cdot \textbf{S}$ with $D=19$ along (a) horizontal bond and (b) diagonal bonds, imposing vertical stripes. (c) Order parameter $\sigma$ measuring $C_{4v}$ symmetry breaking, as defined in Eq.\ref{eq:sigma}. The red (resp black) line marks the first (resp. second) order transition line (see text).
 The vicinity of a critical point can be inferred from the rapidly changing nature of the correlations around $J_2 = 0.37, T=0.1$. In the ferromagnetic region, we completed PEPS data with the exact zero temperature value.}}
\label{fig:sds}
\end{figure*}

We show the bond energy $\ev{\textbf{S}\cdot \textbf{S}}$ in Fig.~\ref{fig:sds}, where we plotted the mean value of this observable over the unit cell for (a) horizontal bonds and (b) diagonal bonds, computed with $D=19$.  ``Horizontal'' and ``vertical'' are the same until lattice rotation is broken, then we redefined them so that stripes always spread along the vertical direction and the order parameter is always positive. In the ferromagnetic phase, far enough from the transition, we added exact zero temperature correlations from the ferromagnetic ground state to complete our numerical data. 
The black dashed line marks the second order transition obtained following the method explained in Fig.~\ref{fig:t2_extra}.

The red points labels the first order transition, where a cubic fit of the free energy cuts the ferromagnetic MSWT, as explained in Fig.~\ref{fig:mswt}. The upper value for the error bar is determined by a linear fit of the free energy. The red dotted line is an artist view of the first order transition line, making use of all the information we gathered. It starts from the zero temperature point at $J_2^c = 0.394$ with an infinite slope and a square root behavior as obtained in Eq.~(\ref{eq:sqJ2}), bends to the right before turning to the left and reaches the free energy crossing points. The critical point exact position is not known, we forecast it should be very close to the point at $J_2 = 0.37$ since correlations change nature rapidly in this region. On panel (c), we plot the order parameter $\sigma$ as defined in Eq.~(\ref{eq:sigma}). It acquires a non-zero value only in the stripe phase, below the second order transition. 

\begin{figure}
\centering
\includegraphics[width=1.0\linewidth]{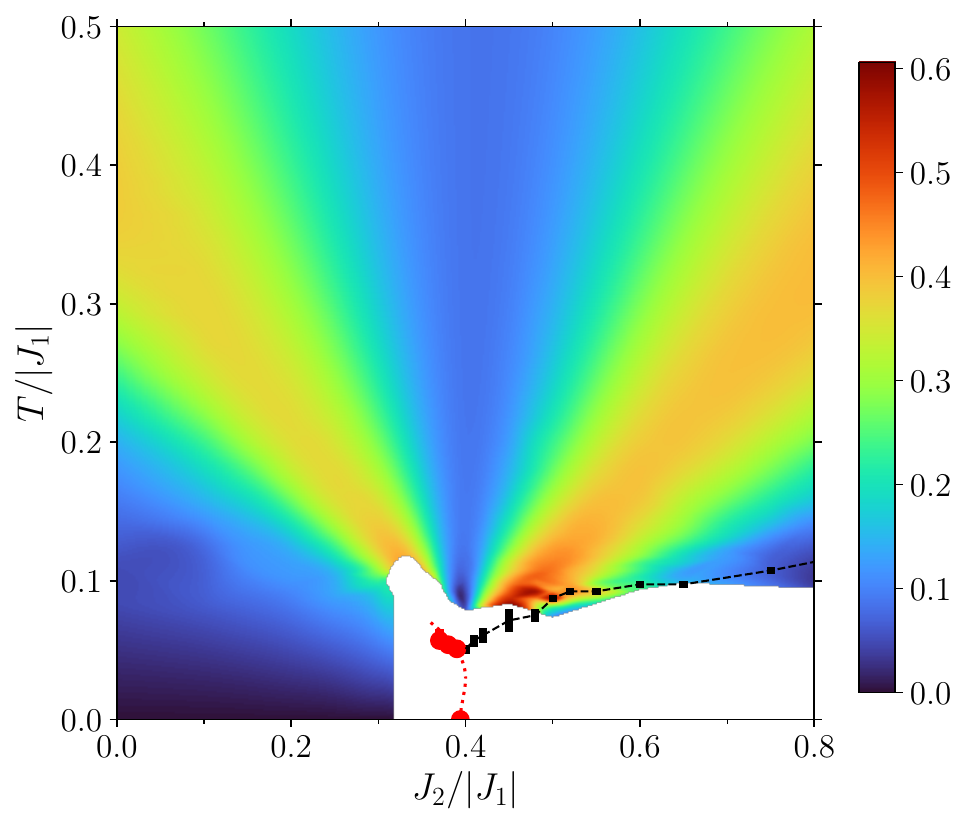}
\caption{\footnotesize{Specific heat for the ferromagnetic-antiferromagnetic $J_1-J_2$ model.  In the ferromagnetic phase, whenever PEPS provided energy values beyond the specific heat maximum we fitted the free energy and used the specific heat Taylor expansion to complete data down to $T=0$.}}
\label{fig:colorplot_C}
\end{figure}

We now look at the specific heat in Fig.~\ref{fig:colorplot_C}. In the ferromagnetic region, we know the exact ground state energy as well as the temperature dependence that obeys $E(T) = E_{\text{FM}} + bT^2 + cT^3 + O(T^4)$. For temperatures below the specific heat maximum, this Taylor development at $T=0$ becomes valid for the specific heat and we used these values to complete our data. This is only possible for $J_2$ values where we could reach the specific heat maximum. 

In the large $J_2$ region, the second order transition is associated with a diverging specific heat.
For very large values of $J_2$, the specific heat peak associated with the Ising transition becomes too narrow to be clearly seen. We can still detect the transition by measuring a rising order parameter and we deduce the presence of the divergence.

Away from the phase transitions, we find two broad specific heat peaks at high temperature, a.k.a. Schottky anomalies above the two different phases. We clearly see on the zoom Fig.~\ref{fig:zoom} that they emerge from two different origins: as presented in the sketch Fig.~\ref{fig:sketch}, the peak above the ferromagnetic phase starts at the critical point around $J_2 = 0.37$. It can be seen as a continuation of the first order line into a crossover and confirms the slope of this line.

Following the large $J_2$ Schottky-like broad peak to lower values of $J_2$ is more complicated. This peak is associated with the building of second neighbor correlations ruled by $J_2$, and we checked that for large $J_2$, its position and width are constant as a function of $T/J_2$. As it is a purely $J_2$ phenomenon, independent from $J_1$, we believe it to exist everywhere above the $J_2$ dominated phase and to be cut by the first order line (see  Fig.~\ref{fig:sketch}). Our numerical calculations with $D=19$ fail to capture this peak for $J_2 = 0.39$ and $J_2 = 0.40$, therefore this peak cannot be seen in this range of $J_2$ in Fig.~\ref{fig:zoom}. For $J_2 = 0.39$, the peak may indeed not exist, however we predict a second order transition for $J_2 = 0.40$ so it should be present for that value of $J_2$. We think this is a finite $D$ effect: larger $D$ values are required to describe fine details around the maximally frustrated region. We observed a similar behavior for $J_2 = 0.41$, where the peak is not seen with $D=16$ but starts appearing with $D=19$.


\section{Conclusion}
To summarize, we used tensor networks to study the spin-1/2 $J_1-J_2$ Heisenberg model with ferromagnetic $J_1 < 0$ and antiferromagnetic $J_2 > 0$ on the square lattice at finite temperature. To circumvent convergence issues, we combined PEPS and spin waves analysis to localize the first order line out of the ferromagnetic phase. We found the first order line has a complex shape, with an infinite slope at $T=0$, a bending to the right before a U-turn towards lower $J_2$.
We also evaluated the critical  temperature for the second order phase transition at large $J_2$ associated with the appearance of stripes breaking lattice rotation symmetry. In the intermediate region, we found that this second order line terminates when it meets the first order line with a critical end point around $J_2 = 0.39$. We found no evidence for an intermediate phase in the model, in agreement with Ref.~\cite{jiang_where_2023}.

This work highlights the remarkable consequences of quantum fluctuations, as they generate sharp differences from either classical or antiferromagnetic $J_1 > 0$ cases. 
Its natural extension would be to study the effect of an external field and the possibility to observe a finite temperature transition to the intermediate, spin nematic phase. We leave this for future work.


\begin{acknowledgments}
 We thank Pratyay Ghosh for indicating us relevant literature. We are grateful to Loïc Herviou and Mithilesh Nayak for insightful discussions.

The PEPS computations have been performed on the facilities of the
Scientific IT and Application Support Center of EPFL.
The source code is available from the corresponding author upon reasonable request. The analysis developed here can be reproduced using raw PEPS data and scripts stored on Zenodo at the address \url{https://zenodo.org/record/8436526}.

The Flatiron Institute is a division of the Simons Foundation.

\end{acknowledgments}

\bibliography{../thermal_ferroJ1_J2}

\appendix

\section{Symmetry implementation}
\label{ap:sym}
We implemented the full SU(2) symmetry in the simple update, however performance limitations prevented us from using it for CTMRG. We improved the usual U(1) implementation by combining it with $S^z$ reversal. This operation noted $Z$ in the following does not commute with U(1) rotation, the subgroup of SU(2) including all these operations is the semi-direct product group $\mathrm{U}(1) \rtimes \mathbb{Z}_2$, isomorphic to the orthogonal group O(2).

The  group O(2) is non-abelian, fortunately its representation theory can be easily recovered by looking at the effect of $S^Z$ reversal on U(1) irreducible representations (irrep). We recall that U(1) is abelian, its irreps are all 1-dimensional and labeled by an integer $n \in \mathbb{Z}$. When U(1) is understood as a subgroup of SU(2), we have the identification $n = 2S^z$. As $S^z$ reversal naturally corresponds to flipping the sign of $S^z$, we expect it to unify the U(1) irreps $\ket{+n}$ and $\ket{-n}$ into a 2-dimensional O(2) irrep $\ket{\pm n}$. There are in practice three cases to consider:
\begin{itemize}
\item  For even $n=2m \neq 0$ (integer $S^z$), this is a linear representation $Z \ket{2m} = \ket{-2m}$.
\item  For odd $n=2m+1$ (half-integer $S^z$), this is a projective representation as $Z$ adds a sign only from negative to positive: assuming $m \geq 0$, we have $Z \ket{2m+1} = \ket{-2m-1}$ but $Z \ket{-2m-1} = -\ket{2m+1}$.
\item For $n=0$, $Z$ maps the vector $\ket{n}$ to another one in the same symmetry sector, in other words $Z$ commutes with $S^z$ in this subspace. Therefore we can define two different cases that are even and odd according to the operation: $Z \ket{0_e} =\ket{0_e} $ (corresponding to the $S^z = 0$ state of even integer spins) and $Z \ket{0_o} = - \ket{0_o} $ (corresponding to the $S^z = 0$ state of odd integer spins).
\end{itemize}
Note that for projective representations, our convention differs from the standard SU(2) one, with a negative sign under $S^z$ reversal always from negative to positive $S^z$ (for SU(2), this depends on the total spin).

In this language, the group fusion rules are written
\begin{eqnarray*}
&0_e \otimes 0_e = 0_e  \\                                                               
&0_e \otimes 0_o = 0_o  \\                                                               
&0_o \otimes 0_o = 0_e  \\                                                               
&0_e \otimes \pm n = \pm n  \\                                                       
&0_o \otimes \pm n  = \pm n   \\                                                       
&\pm n  \otimes \pm n  = 0_e \oplus 0_o \oplus \pm 2n  \\                            
&\pm n  \otimes \pm n  = \pm (n-m) \oplus \pm (n+m) 
\end{eqnarray*}

We recall that to implement U(1) in tensor networks, U(1) quantum numbers are associated to every leg. Any tensor acquires a block-diagonal structure and an efficient implementation only stores these blocks with a well defined symmetry sector.

To upgrade this implementation to O(2), we now use O(2) quantum numbers, merging together $\pm n$ labels. We keep the U(1) blocks with a U(1) label $S^z > 0$ and use them as a representative for the O(2) $\pm S^z$ block. We discard all symmetry blocks corresponding to $S^z < 0$ as they can be recovered by applying $S^z$ reversal. More importantly, we split the $S^z=0$ sector into 0-even and 0-odd. Any tensor operation (contraction, SVD, \dots) can be proceeded using these symmetry blocks.

In terms of performances, we automatically gain a factor 2 in memory as we store twice less data. We also obtain a surprisingly large speed-up. In the case of SU(2)-symmetric problems, the bottleneck always lies in operations in the $S^z = 0$ sector that combines all integer spin values. In the case of finite temperature with purification, the virtual space contains only integer spins therefore all total spin values appear in this U(1) sector. With O(2) implementation, this block is split more or less equally into even and odds sectors: for cubic complexity operations such as contraction or SVD, this gives a factor 8 in time! In practice the bottleneck is moved to the $S^z = 1$ sector, whose size did not change, although we still gain a factor 2 by not dealing with the  $S^z = -1$ block.

\section{Modified spin wave theory}
\label{ap:mswt}
We build on Takahashi's modified spin wave theory (MSWT)~\cite{takahashi_quantum_1986, takahashi_few-dimensional_1987} for a Heisenberg ferromagnet with first neighbor interaction $J_1$ and add a second neighbor $J_2$ interaction. In this framework, we consider a finite size system of $N$ spins with length $S$ and introduce a chemical potential $\mu$ to control the number of magnons and preserve $\expval{S^z} = 0$ at any finite temperature.
Taking into account $J_2$, the magnon energy becomes
\begin{align}
\epsilon_\textbf{k} = & 4 S \abs{J_1}\left(\sin^2\frac{k_x}{2} + \sin^2\frac{k_y}{2}\right) \nonumber \\
 - & 4SJ_2\left(\sin^2\frac{k_x + k_y} {2} + \sin^2\frac{k_x-k_y}{2}\right)
\end{align}

Our basis is the set of occupation  magnon number states
\begin{equation}
\ket{\{n_\textbf{k}\}} = \prod_\textbf{k} (n_\textbf{k}!)^{-1/2} (a^\dagger_\textbf{k})^n_\textbf{k} \ket{0}
\end{equation}
in which we can express the finite temperature density matrix
\begin{equation}
\rho = \sum_{\{n_\textbf{k}\}} \prod_\textbf{k} P_\textbf{k}(n_\textbf{k})\ket{\{n_\textbf{k}\}}\bra{\{n_\textbf{k}\}}
\end{equation}

The $ P_\textbf{k}(n_\textbf{k})$ are probabilities that obey the constraint
\begin{equation}
\forall  \textbf{k} \quad \sum_{n=0}^\infty  P_\textbf{k}(n) = 1,
\label{eq:proba}
\end{equation}
which is associated with a Lagrange multiplier $\mu_\textbf{k}$.

The thermodynamic entropy is given by
\begin{equation}
\text{entropy} = -\sum_\textbf{k} \sum_{n=0}^\infty  P_\textbf{k}(n) \ln P_\textbf{k}(n).
\end{equation}

It allows us to compute the free  energy
\begin{equation}
F_0 = \sum_\textbf{k} \epsilon_\textbf{k} \sum_{n=0}^\infty  n P_\textbf{k}(n) + T \sum_\textbf{k} \sum_{n=0}^\infty  P_\textbf{k}(n) \ln P_\textbf{k}(n_\textbf{k})
\end{equation}

Since SU(2) cannot be broken, we impose the constraint of zero magnetization along $z$-axis, i.e. $S$ magnons per site:
\begin{equation}
0 = NS - \sum_{\vb{k}} \sum_{n=0}^\infty n  P_{\vb{k}}(n). \label{eq:part_number}
\end{equation}
This constraint on the particle number introduces another Lagrange multiplier that is by definition the chemical potential $\mu$. Therefore we have to minimize
\begin{equation}
W = F_0 -  \sum_{\vb{k}} \mu_{\vb{k}}  \sum_{n=0}^\infty n  P_{\vb{k}}(n) - \mu \sum_{\vb{k}}  \sum_{n=0}^\infty n  P_{\vb{k}}(n)
\end{equation}

Imposing a vanishing derivative along $P_{\vb{k}}(n)$ and using the constraint on probability sum (\ref{eq:proba}) to get the first expression for $P_{\vb{k}}(n)$ gives
\begin{equation}
P_{\vb{k}}(n) = (1 - \exp\alpha_{\vb{k}} ) \exp(n\alpha_{\vb{k}}),
\end{equation}
where we introduced $\alpha_{\vb{k}} =(\mu - \epsilon_{\vb{k}}) / T$.
\\

We now consider the constraint on the number of particles (\ref{eq:part_number}). We define
\begin{equation}
\tilde{n}_{\vb{k}} =  \sum_{n=0}^\infty n  P_{\vb{k}}(n) = \frac{1}{e^{-\alpha_{\vb{k}}} - 1}
\end{equation}
and we obtain the self-consistency equation for the chemical potential
\begin{equation}
NS = \sum_{\vb{k}} \tilde{n}_{\vb{k}} =  \sum_{\vb{k}} \frac{1}{e^{-\alpha_{\vb{k}}} - 1}
\label{eq:SN}
\end{equation}
which is nothing but Bose-Einstein statistics.
We can now insert (\ref{eq:SN}) in the free energy $F_0$:
\begin{equation}
F_0 = \mu N S - T \sum_{\vb{k}} \ln (1 + \tilde{n}_{\vb{k}})
\label{eq:F0}
\end{equation}

In his original work, Takahashi computes $\expval{\vb{S}_i\cdot\vb{S}_j}$ to obtain a higher order correction of the free energy. We limited our computation to the first order and observed negligible deviation from his values.
Consequently we only consider the linear spin wave Hamiltonian
\begin{equation}
H_0 = S  \sum_\textbf{k} \epsilon_\textbf{k} a^\dagger_\textbf{k} a_\textbf{k}
\end{equation}
and neglect higher order corrections. The self-consistency equation (\ref{eq:SN}) can be solved numerically to obtain the chemical potential $\mu$ at any temperature. In our case, we have $J_1 = -1$, $S=1/2$ and we recover the implicit equation on $\mu$ from Eq. (\ref{eq:mu})
\begin{equation}
\frac{1}{2\pi^2}\iint_{\text{BZ}} \frac{\dd^2 \vb{k}}{\exp(\beta(\epsilon_{\vb{k}} - \mu)) - 1} - 1 = 0
\end{equation}

In practice, we make a change of variable $(x, y) = \textbf{k}/2$ and use $\epsilon_\textbf{k}$ symmetries over the Brillouin zone to write the more numerically stable form
\begin{equation}
\int_{x=0}^\frac{\pi}{2}\int_{y=0}^{x} \frac{\dd x \dd y}{\exp(\beta(\epsilon_{x,y} - \mu)) - 1} - \frac{\pi^2}{16} = 0.
\end{equation}

Finally the free energy per site is obtained from (\ref{eq:F0}) as
\begin{equation}
f_0 = E_{\text{FM}} + \frac{\mu}{2} - \frac{1}{4\pi^2\beta}\iint_{\text{BZ}} \ln(1 + \frac{1}{\exp(\beta(\epsilon_{\vb{k}} - \mu))-1})\dd^2 \vb{k}
\end{equation}
with $E_{\text{FM}} = -1/2 + J_2 / 2$.

\section{PEPS parameters control}
\label{ap:control}
Our PEPS algorithm is controlled by three parameters: the imaginary time step $\tau$,  the virtual bond dimension $D$, and the corner dimension $\chi$. They become exact in the limits $D\rightarrow \infty$, $\tau \rightarrow 0 $ and $\chi \rightarrow \infty$. In the current case, $D$ is by far the most important parameter.

Most of our simulation are run with $\tau = 10^{-3}$, we checked that changing it within a range $5 \cdot 10^{-5} \leq \tau \leq 5 \cdot 10^{-5} $ has no visible effect on any observables. It can however affects CTMRG convergence: stability is improved by reducing the lattice asymmetry induced by the simple update, and smaller $\tau$ offers better chances to converge.

We typically use a corner dimension $\chi = D^2$, but for $D=19$ even smaller values are enough to converge local observables away from phase transitions.
In the critical region, very close to the second order phase transition, effects are stronger, especially for the order parameter $\sigma$, but convergence can still be reached as can be seen in Fig.~\ref{fig:t2_extra} (b). Transfer matrix spectra and correlation lengths are more sensitive, which leads to less precise values for the critical temperature $T_2$ obtained by gap extrapolation.

\begin{figure}[ht]
\includegraphics[width=1.0\linewidth]{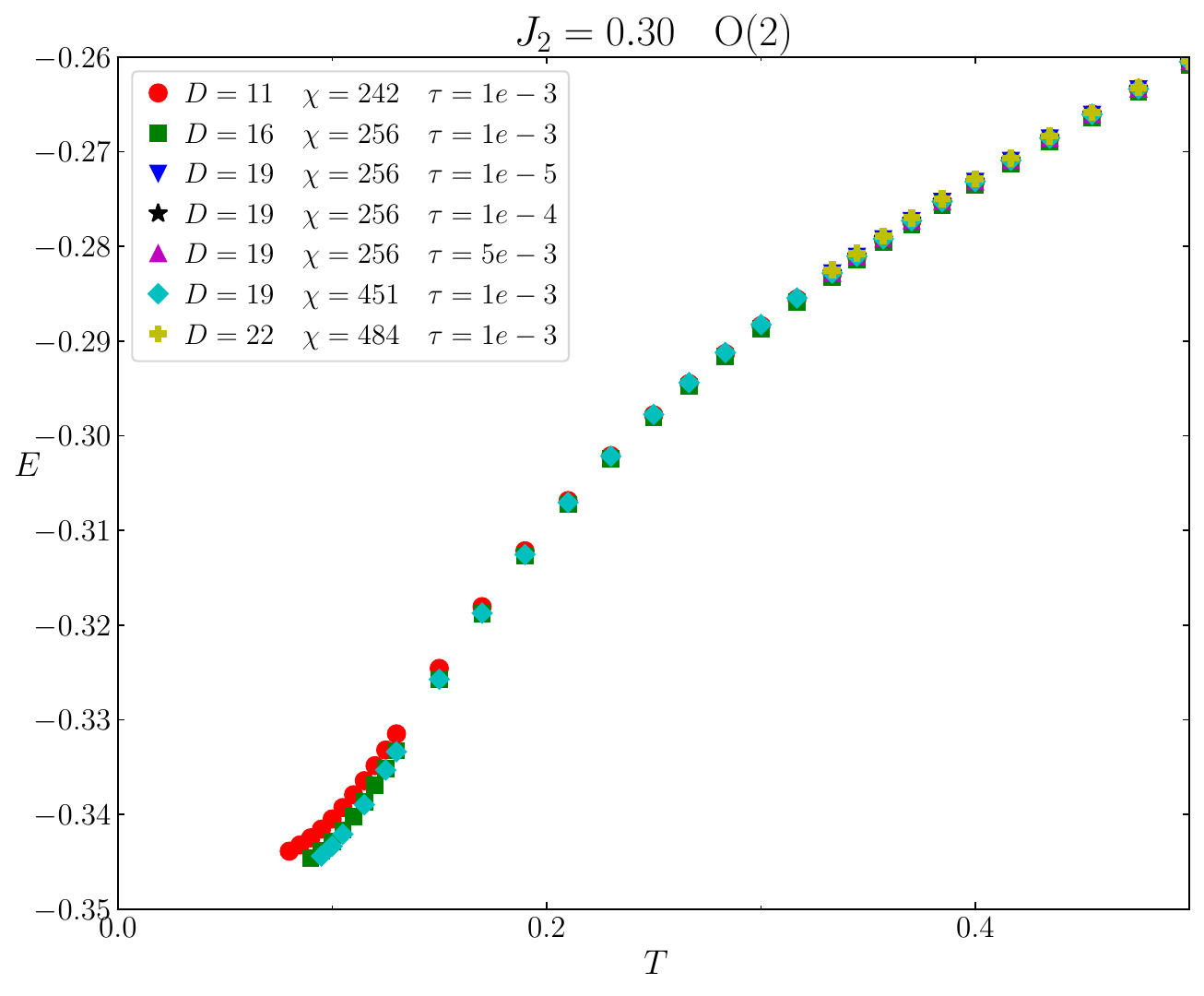}
\caption{\footnotesize{Energy benchmark for different control parameters at $J_2 = 0.30$. In the ferromagnetic phase, quantum effects play a minor role and finite $D$ effects stay small. $\tau$ effects are always negligible.}}
\label{fig:benchmark030}
\end{figure}

The bond dimension $D$ plays a more important role. In the ferromagnetic phase, the quasi-classical behavior allows to obtain good results even for moderate values of $D$, as we show in Fig.~\ref{fig:benchmark030}. In the antiferromagnetic region, $D$ effects are much stronger: the Ising phase transition only appears for $D \geq 16$. In the intermediate region $J_2 \sim 0.40$, $D$ also affects the presence of the Schottky anomaly. For larger $J_2$, changing $D$ from 16 to 19 slightly changes the critical temperature $T_2$ as we already discussed in the $J_1 > 0$ case~\cite{gauthe_thermal_2022}, but it does not alter the qualitative behavior.

\end{document}